\documentstyle[aps,preprint,tighten,floats,epsf,rotate]{revtex}

\begin{document}
\draft
\title{Stochastic Production Of Kink-Antikink Pairs In The Presence Of An 
Oscillating Background}
\author{Rajarshi Ray$^{a}$ \footnote{rajarshi@iopb.res.in} and 
Supratim Sengupta$^{b}$ \footnote{sengupta@phys.ualberta.ca}}
\address{\it $^a$Institute of Physics, Sachivalaya Marg, Bhubaneswar 751005, 
India. \\
$^b$Theoretical Physics Institute, Department of Physics, University of 
Alberta, Edmonton, Canada T6G 2J1.}
\date{November 2001}
\maketitle
\widetext
\parshape=1 0.75in 5.5in
\begin{abstract}

 We numerically investigate the production of kink-antikink pairs in a 
$(1+1)$ dimensional $\phi^4$ field theory subject to white noise and 
periodic driving. The twin effects of noise and periodic driving acting 
in conjunction lead to considerable enhancement in the kink density
compared to the thermal equilibrium value, for low dissipation coefficients
and for a specific range of frequencies of the oscillating background. The 
dependence of the kink-density on the temperature of the heat bath, the 
amplitude of the oscillating background and value of the dissipation 
coefficient is also investigated. An interesting feature of our result is 
that kink-antikink production occurs even though the system always remains 
in the broken symmetry phase.  

\end{abstract}
\vskip 0.125 in
\parshape=1 -.75in 5.5in
\narrowtext
 
\section{Introduction}

  During the past decade, the evolution of quantum and classical fields 
out of equilibrium has received a lot of attention. The interest generated
in this subject matter is motivated by many physical scenarios in the early
universe as well as in condensed matter systems, where non-equilibrium 
phenomenon plays a crucial role. Of particular relevance is the formation
and evolution of quark-gluon plasma, the formation of topological defects,
out-of-equilibrium phase transition dynamics and its impact on domain 
formation and growth. The non-equilibrium evolution of fields is also a
crucial input in understanding how the current baryon asymmetry of the 
universe was generated. 

  The formation of topological defects \cite{vil} is a generic feature of 
most symmetry breaking phase transitions which give rise to a vacuum manifold 
with a non-trivial topology. The most widely discussed mechanism of 
topological defect formation during phase transitions was first put forward
by Kibble in a seminal paper twenty-five years ago \cite{kibble}. A crucial
feature of this mechanism is that it depends solely on the topology of the
vacuum manifold and on the space-time dimension, but not on the details 
of the field dynamics, which makes
it universally applicable across a diverse range of energy scales. This 
also allows many of the predictions of defect formation in the early universe
to be tested in the laboratory using condensed matter systems like superfluid
Helium and liquid crystals \cite{zurek}. Recent experiments with liquid
crystals have spectacularly demonstrated the validity of Kibble mechanism
in first order phase transitions by testing universal aspects of the 
predictions \cite{ajit1,ajit2}. A complete understanding of defect formation 
in systems (like superfluid Helium-IV) undergoing second order phase 
transitions is still lacking \cite{rivers}. 

Dynamics can and does play a crucial role in topological defect formation. 
It has been demonstrated \cite{tanmay}, that the density of defects in 
first order transitions depends on the velocity of the bubble wall which is 
determined by the dissipation coefficient of the medium. Moreover, the 
dynamics of bubble collisions in a first order transition has been shown to 
lead to a new mechanism of defect formation stemming from the flipping of the 
order parameter field through the zero of the field. For the case of a 
spontaneously broken global U(1) field theory in $(2+1)$ dimensions, this 
results in the {\it discontinuous} change in phase of the field by $\pi$, 
leading to the formation of a vortex-antivortex pair. (In $(3+1)$ dimensions, 
a string loop is produced). This mechanism first observed in systems
where the symmetry is spontaneously as well as explicitly broken \cite{digal}
was later shown to be generically valid and investigated in detail \cite{ssg1}. 

In second order phase transitions, the phenomenon of critical slowing down 
\cite{zurek1} (freezing of field dynamics near the critical point) is crucial 
in determining the initial density of defects. The dependence of defect 
density on the quench rate of a second order transition has also 
been extensively investigated \cite{zurek2,rivers2,calz}.

Until recently, conventional wisdom suggested that topological defect 
formation could occur only during phase transitions. However, it has been 
recently shown \cite{ray} that topological defects can also be produced by 
the flipping mechanism \cite{ssg1} under conditions in which the system always 
remains in the broken symmetry phase without undergoing any phase transition. 
The production of vortex-antivortex pairs was found to be induced by field 
oscillations brought about by the coupling of the U(1) scalar field to a 
periodically oscillating background \cite{ray}. The flipping of the 
order-parameter field in localized regions resulted in the  formation of 
vortex-antivortex pairs, for a certain range of resonant frequencies of 
the periodically oscillating background.  

 It is then natural to ask how the defect densities would be affected in 
the presence of noise. After all, most physical systems in nature are 
not closed but are continually involved in exchanging energy 
with a heat bath \cite{boyan,hu}. The heat bath often represents other 
degrees of freedom which interact with the system in accordance with
the fluctuation-dissipation theorem. In this paper we address this issue 
by studying the formation of kink-antikink pairs in $(1+1)$ dimensions.

 We should also mention that a Langevin dynamics approach has also been 
used to study bubble dynamics in a noisy background \cite{abney},long lived
oscillating field configurations (oscillons) in a thermal bath \cite{haas}
and to obtain the nucleation rates of bubble formation in a first order 
phase transition using both additive and multiplicative noise \cite{suraj}.  

 The investigation of kink-antikink production has been carried out in a 
variety of contexts. The non-equilibrium dynamics of kink formation was 
investigated \cite{antunes} for a damped scalar field theory undergoing a 
symmetry breaking phase transition with the aim of understanding how the 
defect densities and correlation lengths depend upon the choice of 
dissipation coefficient and initial conditions. The regime of validity of 
the linear and Hartree approximation was also elucidated and it was shown 
that non-linear effects play a crucial role in determining the initial 
defect densities after a quench. The thermal production of kinks has 
been investigated both analytically and numerically. Scalapino, Sears 
and Ferrell used the transfer integral 
technique to calculate the exact partition function and correlation functions 
for a dilute gas of kink-antikink pairs \cite{ssf}. Krumhansl and Schreiffer 
(KS) later showed that in the low temperature limit where the dilute gas 
approximation is valid, the kink contribution to the partition fuction
can be factored out and identified with a tunneling term \cite{ks}. Currie
et.al. \cite{ckbt} further generalized the results of KS by taking into 
account the interaction of the kinks with phonons. They also studied kink 
production in the Sine-Gordon model in addition to the real scalar field model. 
Numerical investigations to check the theoretical predictions like the thermal
nucleation rate and kink life-time have also been carried out \cite{grig,glsr1}. 
More recently, Alexander, Habib and Kovner \cite{habib} investigated thermal 
production of kinks both analytically and numerically by using a Langevin 
equation with additive white noise to model the effect of thermal 
fluctuations. They were able to identify the temperature regime 
below which the dilute gas (WKB) approximation is valid. By introducing a 
double-gaussian approximation they were able to obtain an excellent 
agreement between their theoretical prediction of thermal kink densities 
and their numerical results. Langevin simulations in the intermediate 
temperature range were recently carried out by Gleiser and Muller 
\cite{glsr2}, where they also pointed out the important issue of lattice 
spacing dependence of results in simulations of stochastic field equations.
The thermal nucleation of interacting kink-antikink pairs in the Sine-Gordon
model has been investigated by B\"{u}uttiker and Landauer, in the overdamped
limit \cite{buttiker} . The nucleation rate in the overdamped
limit was found to be proportional to the square of the equilibrium density
\cite{buttiker}; however some studies also suggested \cite{hanggi} that the
nucleation rate is proportional to the cube of the equilibrium density of
kinks. The resolution of this controversy depends on unambiguously ascertaining
whether the kink lifetime is inversely proportional to the equilibrium kink
density \cite{christen} or the square of the equilibrium kink density
\cite{hanggi}. Recent work for both Sine-Gordon \cite{christen} and $\phi^4$
models \cite{lythe2}, invloving extensive numerical simulations of the
nucleation and annihilation dynamics of thermal kinks and antikinks, have
clarified that the nucleation rate is proportional to the square of the
equilibrium density. In all these papers, the effect of an oscillating
background driving the kink-producing field was not considered. Such a
situation has been discussed by Marchesoni et.al. \cite{march} for an
overdamped field theory.
 
 In this paper we focus on numerically investigating the formation of 
kinks in a $(1+1)$ dimensional spontaneously broken {\it relativistic} 
(underdamped) scalar field theory coupled to an oscillating background 
and subject to white noise. \footnote{Strictly speaking there is no phase 
transition in $d \leq 2$ spatial dimensions, for systems in equilibrium 
\cite{mermin}. However, in our case, the 
system is always evolving out-of-equilibrium and so the Mermin-Wagner 
theorem does not apply.} The coupling of the field to the oscillating 
background and thermal noise induces large amplitude field oscillations 
for a certain range of frequencies, thus enabling the field to cross over 
the potential barrier resulting in the nucleation of kink-antikink pairs. 
We find that kink-antikink pairs are produced inspite of the fact that the 
system both initially and for all subsequent times, remains in the broken 
symmetry phase. This has important consequences for topological defect 
production \cite{ray} since it implies that topological 
defects need not be produced only during a phase transition, as was 
believed earlier. Moreover, kink-antikink production is found to depend 
sensitively on the dissipation coefficient and the effect becomes 
considerably suppressed for high dissipation coefficients. 

 The physical situation we discuss is of relevance to both early universe 
as well as condensed matter physics. During reheating after 
inflation, the inflaton field starts oscillating about its vaccum value and
would act like a driving force to any other scalar field to which it is 
coupled. The oscillations eventually die out due to particle production
caused by transfer of energy from the oscillating inflaton to the quanta 
of fields to which it is coupled, leading to reheating of the universe 
and a transition from matter dominated to the radiation dominated phase. 
The process of (p)reheating of the universe after inflation has been the 
subject of intensive study during the last decade during which a new theory 
of reheating (due to inflaton decay via explosive soft particle production)
was developed and studied in detail \cite{in1}. The effect of noise on the 
growth of fluctuations has also been studied \cite{in5} in the context of 
reheating after inflation. 
Domain wall \cite{parry}
and cosmic string \cite{in3} production during (p)reheating has also been 
investigated. The main premise of these papers was that (p)reheating could 
result in non-thermal symmetry restoration \cite{in11} after inflation. 
Topological defects would then be produced during the subsequent symmetry 
breaking brought about by rescattering effects and/or cooling due to 
expansion, in the usual manner (i.e. via Kibble mechanism). However, as 
pointed out recently \cite{ray}, topological defects can form even 
{\it without} the system undergoing any thermal or non-thermal phase 
transition; simply because of large amplitude oscillations of the defect 
producing field, induced by its coupling to a spatially homogeneous, 
oscillating, inflaton field. 

 In condensed matter systems, the oscillating background can be thought of 
as an external oscillating influence such as temperature, pressure or 
even an electric or magnetic field coupled to the system. A system subject 
to noise and periodic driving has also been extensively studied in non-linear 
dynamics , in the context of stochastic resonance \cite{stch}. The field 
theory system considered here, shares the characteristic features of systems 
undergoing stochastic resonance. However, the phenomenon we observe, i.e. 
enhancement in kink-antikink densities due to the twin effects of noise and 
periodic driving, is distinct from stochastic resonance, as will be discussed 
later.  

  The main purpose of this work is to investigate the effect of noise and 
an oscillating background on the density of kink-antikink pairs. The
presence of the oscillating background ensures that the system is always
out-of-equilibrium and so the defect production in this case is distinct
from the thermal production of kink-antikink pairs 
\cite{ssf,ks,ckbt,grig,glsr1,habib,glsr2,march}. The main result of this
work is the observed enhancement of kink-antikink density, compared
to the thermal equilibrium value, as a result of the twin effects of noise 
and coupling of the field with an oscillating background. 

  The paper is organized as follows. In section II, we describe our 
model and the numerical algorithm we use to solve the Langevin equation. 
There we also discuss some of the issues related to the lattice-spacing
dependence of the results of Langevin simulations. The results of our 
numerical simulations are described in section III. The dependence of 
kink-antikink density on parameters such as bath temperature, amplitude 
of oscillation of the background homogeneous field and the dissipation 
coefficient are investigated. The range of frequencies for which defect 
production occurs is also obtained. We end with a brief summary and 
discussion of our results in section IV. 

\section{The Model and Numerical Techniques}

 The Lagrangian density for a spontaneously broken real scalar field theory
in $(1+1)$ dimensions is  

\begin{equation}
{\cal L} = \frac{1}{2}(\partial_{\mu}\varphi)(\partial^{\mu}\varphi) 
- \frac{\lambda}{4}(\varphi^2 - \varphi_0^2)^2 - \frac{1}{2} g^2 \chi^2 
\varphi^2 
\end{equation}

 where $\varphi$ is a real scalar field coupled to a spatially homogeneous, 
oscillating background field $\chi$ given by $\chi = \chi_0 sin(\omega t)$ 
and $g^2$ is the coupling parameter. In the absence of the coupling term 
($g^2 = 0$), the above Lagrangian density leads to the field equation 

\begin{equation}
\partial^2 \varphi/\partial t^2  - \bigtriangledown^2 \varphi 
+ \lambda \varphi (\varphi^2 - \varphi_0^2) = 0 , 
\end{equation}

whose static solution admits extended but localized topological structures 
called kinks and  antkinks   

\begin{equation}
\varphi(x)_{\pm} = \varphi_{0} tanh[\sqrt{\frac{\lambda}{2}}\phi_0 (x \pm x_0)]
\end{equation}

where the $\pm$ signs correspond to kink and anti-kink located at $x_0$ and
$-x_0$ respectively. It is often convenient to work with dimensionless 
quantities obtained by scaling the variables as follows 

\begin{eqnarray}
\varphi & \rightarrow & \phi = \frac{\varphi}{\varphi_0}, 
\chi \rightarrow \frac{\chi}{\varphi_0}, \nonumber \\
x  & \rightarrow & \sqrt{\lambda} \varphi_0 x,
t \rightarrow \sqrt{\lambda} \varphi_0 t, \nonumber \\
g^2 & \rightarrow & \frac{g^2}{\lambda}  \\
T & \rightarrow & \frac{T}{\sqrt{\lambda}\varphi_0^3}, \nonumber \\ 
\eta & \rightarrow & \frac{\eta}{\sqrt{\lambda}\varphi_0} \nonumber 
\end{eqnarray}

  In terms of the above dimensionless quantities, the kink solution (eq.(3))
is now given by  $\phi(x)_{\pm} = tanh[(x \pm x_0)/\sqrt{2}]$ and 
the kink mass is easily calculated to be $M_{k} = \sqrt{8/9}$. 

 To take into account the effect of interaction of the system with a 
background thermal bath, it is customary to use a stochastic description 
in which the system interacts with a thermal bath at a given temperature $T$. 
The presence of a thermal bath with which the system interacts can be easily 
motivated. Most realistic physical scenarios involve a system exchanging 
energy with a heat bath which may either represent external thermalized 
degrees of freedom  or, in some cases, certain degrees of freedomi, typically
the hard modes, of the system, which thermalize on a shorter time-scale 
compared with the rest of the system \cite{boyan,hu}. These can therefore be 
considered as an environment with which the system, consisting of the 
non-thermal degrees of freedom (those with longer thermalization time-scales; 
usually the soft modes in a field theory), exchange energy. This interaction 
would eventually drive the system to thermal equilibrium (in the absence of 
the oscillating background field). In this description, the Langevin field 
equation, in terms of scaled dimensionless variables, takes the form

\begin{equation}
\frac{\partial^2 \phi}{\partial t^2} + \eta \frac{\partial\phi}{\partial t} - 
\frac{\partial^2 \phi}{\partial x^2} + \phi (\phi^2 - 1) + g^2 \chi^2 \phi 
= \xi(x,t) 
\end{equation}
 
  where the dimensionless damping coefficient $\eta$ and the Gaussian 
(delta correlated) white noise term $\xi(x,t)$ are related by the 
fluctuation dissipation theorem

\begin{equation}
<\xi(x,t)\xi(x^{\prime},t^{\prime})> = 2T\eta 
\delta(x - x^{\prime})\delta(t - t^{\prime})
\end{equation}

to ensure the equilibration of the system in the absence of the coupling 
term. $T$ is the rescaled, dimensionless temperature of the heat bath.

 When the coupling term is small, i.e. for $g^2 \chi_0^2 << 1$, its 
presence does not affect the above form of the kink-antikink solutions. 
The noise and the perturbation term cause the center of mass of the 
kink/antikink to become a random variable \cite{march2}. The small
coupling term also acts as a positive mass term and is responsible for
shifting the vacuum states towards zero. The critical value of $g^2 \chi_0^2$
for which the broken symmetry is restored is $(g^2 \chi_0^2)_{c} = 1$.
To ensure that the system always remains in the broken phase, one
must have 

\begin{equation}
g^2 \chi_0^2 << 1
\end{equation}

 Since the coupling term in the Lagrangian is quadratic in $\phi$, the 
shape of the effective potential remains unchanged. The oscillating background 
field has the effect of modulating the barrier height and the position of the 
two disconnected vacuum states of the potential. During one period of 
oscillation of the background field, the two degenerate vacuum states of 
$\phi$ change its value from a minimum of $\pm\sqrt{1 - g^2 \chi_0^2}$ for 
$t = (2n+1)\frac{\pi}{2\omega}; n = 0, \pm 1, \pm 2, ...$ (corresponding to 
the minimum barrier height separating the two disconnected vacua) to a 
maximum of $\pm 1$ for $t = n\frac{\pi}{\omega}; n = 0, \pm 1, \pm 2, ....$ 
(corresponding to the maximum barrier height). 

 In order to study the production of kink-antikink pairs, due to the effects
of noise and periodic driving, we solve Eq.5 numerically using a stochastic 
second order staggered leapfrog algorithm with periodic boundary conditions. 
Such an algorithm has been previously employed by Gleiser and collaborators 
\cite{glsr1,glsr2,glsr3} to study thermal production of kink-antikink pairs 
as well as to investigate ways of matching Langevin lattice simulation results 
with continuum field theories. We briefly outline the algorithm used and then 
discuss some of the subtleties associated with lattice simulations of 
Langevin equations. 

  The discretised version of Eq.5 using the second order stochastic staggered 
leapfrog algorithm can be cast as 

\begin{equation}
\phi_{i,n+1} = \frac{2 \phi_{i,n} - 
(1 - \frac{\eta \Delta t}{2})\phi_{i,n-1} 
+ (\Delta t)^2(\bigtriangledown^2 \phi -
V^{\prime}(\phi_{i,n}) + \sqrt{\frac{2T\eta}{\Delta t\Delta x}} u_{i,n})}
{1 + \frac{\eta \Delta t}{2}} 
\end{equation}

  
 where `i' and and `n' are the spatial and temporal lattice indices 
respectively. $V^{\prime}(\phi)$ is the derivative of the potential with 
respect to $\phi$ and $u_{i,n}$ is a gaussian random number with zero mean 
and unit variance, i.e. $<u_{i,n}> = 0, <u_{i,n}u_{i,m}> = \delta_{m,n}$
(numerically generated by a Box-Mueller algorithm \cite{teuk}); with 
$u_{i,n}$ being related to $\xi_{i,n}$ by the relation 

\begin{equation}
\xi_{i,n} = \sqrt{\frac{2T\eta}{\Delta t \Delta x}}u_{i,n}  
\end{equation}

 The presence of a lattice introduces natural UV and IR momentum cut-offs 
in the theory since the smallest and the largest momentum modes that can
be probed by the simulation are proportional to inverse lattice size 
($\frac{2 \pi}{L}$) and inverse lattice spacing ($\frac{\pi}{\Delta x}$) 
respectively. Here $L = N \Delta x$ is the lattice size, N being the total 
number of lattice points.  Finite-size effects can be ruled out by using a
large lattice size. However, computational constraints prevent the choice
of an arbitrarily small lattice spacing. Also in our case, the presence of an 
external time scale specified by the frequency of the oscillating background 
field requires 

\begin{equation}
\Delta t << \omega^{-1} 
\end{equation}

  The choice of lattice spacing is also crucial especially in 
Langevin simulations as has been pointed out earlier \cite{glsr2,glsr3,lyth}. 
The results of Langevin simulation turn out to
be lattice-spacing dependent unless appropriate counter-terms are introduced
in the effective potential. 
There is some ambiguity about the manner in which the results scale with 
$\Delta x$. A perturbative counter-term linear in $\Delta x$ when added to 
the potential, was found \cite{glsr2} to be adequate enough to remove the 
lattice spacing dependence of the Langevin simulation results. However, it
has been argued \cite{lyth} that a perturbative method is often inadequate 
in obtaining the correct manner in which the simulation results scale with 
$\Delta x$. In particular, the one-loop perturbative procedure  \cite{glsr2} 
does not give any corrections for the free theory. Using a non-perturbative 
approach based on an exact solution of the thermal partition function, 
Bettencourt et.al. \cite{lyth} were able to show \footnote{In this method, 
the nature of the counter-term depends on the time-stepping algorithm used 
for evolving the Langevin equation. An Euler differencing scheme was used in 
\cite{lyth}} that the convergence of the results with lattice spacing scales 
as $(\Delta x)^2$ and not as $\Delta x$ as the perturbative one-loop 
calculation \cite{glsr2} suggested. In both papers, the systems considered 
were in thermal equilibrium and this aspect facilitated the calculation 
of appropriate counter-terms, perturbatively \cite{glsr2}, and through an 
exact computation of the thermal partition function \cite{lyth}. 

 In our case however, the presence of the coupling (of the field) with the 
oscillating background and the low dissipation coefficient (which makes 
transfer of energy between the heat bath and the system inefficient) prevents 
the thermalization of the system even for large times for which the 
simulations were run. Adding counter-terms calculated for systems in thermal 
equilibrium is therefore not helpful in removing the lattice spacing 
dependence of the results. At this stage, the issue of removal of lattice 
spacing dependence of Langevin simulations for non-equilibrium systems 
remains unresolved. We hope to address this problem in a future publication.

 We have carried out our simulations with the spatial lattice spacing 
$\Delta x = 0.4$, the temporal lattice spacing $\Delta t$ = 0.01. The 
physical lattice size (L) was kept fixed at 6553.6 which corresponds to 
$N = 16384$  lattice points for $\Delta x = 0.4$. We have also set $g^2=1$
in all our simulations so that the condition (7) reduces to 

\begin{equation}
\chi_0^2 << 1 
\end{equation}

 The issue of how to identify kinks in a Langevin lattice simulation has 
attracted much controversy \cite{habib,glsr2}, mainly because of the fact that 
at high temperatures, it becomes extremely difficult to distinguish between 
kinks, phonons and large amplitude, nonperturbative and non-topological 
fluctuations. It is fair to say that an unambiguous technique for counting
kinks on the lattice remains to be discovered. In this paper, we follow the 
technique employed in \cite{habib} where a zero crossing of the order 
parameter field is counted as a kink only if there are no other zero 
crossings for one kink width to its left and right. The kink width in terms
scaled and dimensionless units is $2\sqrt{2}$ which corresponds to approx. 8
lattice units, for $\Delta x=0.4$. The total number of kinks and antikinks 
is just twice the number of kinks counted because kinks and antikinks are 
always produced in pairs. Since in these simulations, we are interested only 
in the low temperature regime, where kinks are easily identifiable as 
shown in Fig.1, this method provides a fairly accurate way of counting kinks.  

\section{Numerical Results}

  We are now in a position to describe the results of our numerical 
simulations. The initial conditions of our simulations correspond to the 
situation in which the field over the entire lattice, undergoes small 
amplitude fluctuations about the positive of the two degenerate vacuum states.
The amplitude of fluctuations are of ${\cal{O}}(10^{-3})$ and 
incapable of taking the field over the potential barrier. We have checked 
that our results are independent of initial conditions and remain unchanged 
even if we choose the initial field configuration to be spatially homogeneous 
over the entire lattice with its value corresponding to either one of the two
degenerate vacuum states. As evident, from our choice of initial conditions, 
there are no kinks/antikinks present initially. The amplitude of oscillation 
of the background field is chosen to be $\leq 0.38$ so that the condition (11) 
is satisfied and this alone is incapable of inducing the field to climb 
over the potential barrier in the absence of coupling to the environment.    

  In the absence of the oscillating background ($g^2=0$), thermal 
fluctuations can make localized portions of the field flip over the potential
barrier, resulting in the formation of a kink-antikink pair 
\cite{habib,glsr2}. However, the density of pairs produced depends on the
strength of thermal fluctuations given by $D = T\eta$ which is a measure of 
the amount of energy transferred from the heat bath to the system. 
The thermal nucleation of kink-antikink pairs is suppressed by the Boltzmann
factor $e^{-M_{k}/T}$. Earlier studies \cite{habib,glsr2} were carried out with 
the noise strength $D \geq {\cal{O}}(10^{-1})$ since their main focus was on 
thermal equilibrium production of kinks. The phenomenon we observe occurs at 
low noise strengths in underdamped systems evolving out of equilibrium. In 
our simulations we restrict the damping coefficient $\eta < 0.05$. The 
temperature of the heat bath is taken to be ${\cal{O}}(10^{-1})$, so the 
noise strength $D \sim 10^{-3}$. For such low noise strengths, the production 
of kinks by thermal fluctuations is few and far between. However, in the 
presence of noise and non-vanishing coupling $g^2$, a dramatic enhancement 
in kink production is observed for a certain range of frequencies of the 
oscillating background field. 

  A periodically modulated non-linear system like the one described by 
Eq.(5) is expected to exhibit resonant behavior for a certain range of 
frequencies of the modulator. The presence of noise acting in conjunction
with the periodic modulation induces large amplitude fluctuations in the
field enabling it (in localized regions) to cross over the potential barrier.
This results in the formation of kink-antikink pairs in a manner similar to
the production of vortex-antivortex pairs discussed recently \cite{ray}. The 
theoretical analysis of this phenomenon is extremely complicated not only 
because of the non-linear nature of a system with infinite degrees of freedom, 
but also because one has to deal with stochastic partial differential 
equations. We have therefore decided to take recourse to numerical simulations 
to investigate this phenomenon. The quantities of importance are the mean field
given by $<\phi(x,t)> \equiv \bar{\phi}(t) = (1/L)\int_{0}^{L}\phi(x,t) dx$,
the fluctuation defined as $\delta \phi(t) = \sqrt{<\phi(x,t)^2> - 
<\phi(x,t)>^2}$ and the density of kink-antikink pairs $n(t) = 2 N_k(t)/L$; 
where $N_k(t)$ is the total number of kinks at a given time, counted in the 
manner described earlier.

To identify the resonant frequency regime, it 
is important to realize that too large an oscillation frequency would cause
the effective potential to change in a time scale which is much smaller than
the destabilization timescale of the field (from its initial state), as 
a result of which the field would not feel the change in the shape of the 
potential. On the other hand, too small an oscillation time scale would 
result in the destabilized field having sufficient time to relax to the 
vacuum state of the changing effective potential (apart from fluctuations
due to the presence of noise). With these considerations in mind we find 
that the range of frequency required to induce resonant amplification of 
field amplitude leading to the enhanced production of kinks is 
$0.3 \le \omega \le 2.5$. 

 We first give results for the variation of the mean field, the fluctuation 
and the kink density with change in the temperature ($T$) of the heat bath.
Fig.2 shows the plots (for a single noise realization) with fixed $\omega, 
\chi_0, \eta$ and $T$ varying from
$0.08 - 0.16$ in steps of 0.02 . The mean field value starting from its 
initial value around 1 decreases to zero and eventually starts oscillating
about zero. On the other hand, the fluctuation grows exponentially in a short
time-scale to its asymptotic value and remains nearly constant thereafter. 
The kink density increases with temperature as expected, the mean field is 
also found to decay more quickly to zero for higher temperatures and the 
fluctuation also grows more steeply signifying an increase in the growth 
exponent. The time around which the fluctuation plateaus out is also the
time around which the average kink-antikink density becomes nearly constant 
implying that kinks and antikinks are nucleated and annihilated at nearly 
the same rate. 
The kink-antikink density for the same set of temperatures but with the 
coupling $g^2 = 0$ (i.e. in the absence of the oscillating background) is
shown in Fig.2(d). A comparison between Fig.2(c) and Fig.2(d) clearly 
shows considerable (by at least an order of magnitude) enhancement in the 
kink density when {\it both} noise and periodic driving is present. This is 
especially evident for low temperatures. The presence of the oscillating 
background as well as the low dissipation coefficient prevents the 
thermalization of the system for time-scales upto which the simulations
were run. Hence, kink-antikink pair production occurs out-of-equilibrium.

 The decay of the mean field value to zero is {\it not} an indication of
symmetry restoration (a common misconception existing in the literature
\cite{parry}) but is indicative of the formation of a large number
of kink anti-kink pairs. To establish this unambiguously we plotted the 
probability distribution of the field over the entire lattice by 
appropriately binning the field values. Fig.3(a) shows the initial 
probability distribution of the field for the entire lattice. In view, of
the choice of initial conditions, the sharp peak in the probability 
distribution about $\phi = 1$ is easily understandable. Fig. 3(b) shows
the field probability distribution at a later time ($t=3000$) after the 
fluctuation has flattened out and the average kink density has become nearly 
constant. The fact that the probability distribution is still peaked about 
the non-vanishing vacuum expectation values clearly implies that the 
symmetry remains broken. In contrast to Fig.3(a) two distinct peaks of 
nearly the same height about $\phi \simeq \pm 1$ are now observed. This can 
be easily explained by the fact that the presence of a large number of 
kink antikink pairs causes the fraction of the field around 
$\phi \sim 1$ to be nearly the same as the fraction around $\phi \sim -1$. 
The generic form of the probability distribution function depicted in 
Fig.3(b) is observed till the end of the simulation, which allows us to
conclude that kink-antikink pairs are produced even though the system 
{\it always} remains in the broken phase (just as in \cite{ray}). This is 
contrary to the situation discussed in the context of topological defect 
production during inflationary (p)reheating \cite{parry,in3}. There, the 
fluctuations grow large enough to restore symmetry and defects are produced 
in the conventional manner when the symmetry is subsequently broken due to 
mode-scattering effects and/or expansion of the universe.

  Fig.4 shows the mean field value, fluctuation and kink density obtained
after averaging over 100 different noise realizations. The upper and lower 
curves correspond to the $\pm 1 \sigma$ standard deviation from the noise 
averaged middle curve. The generic features observed in Fig.2 are also 
seen here. At late time, the $\pm 1 \sigma$ error becomes quite small as 
evident from the fluctuation plots in Fig.4(a). 

  The dependence of the kink density on the amplitude of oscillation 
$\chi_0$ is shown in Fig.5. As mentioned earlier, the coupling term acts 
like a positive mass term and this dictates the choice of $\chi_0$ in 
accordance to constraint (11). An increase in $\chi_0$ does lead to an 
increase in defect density as is evident from Fig.5. However, since our 
main interest lies in studying defect production dynamics for small 
amplitudes of the oscillating background, such that the coupling to the 
oscillating background by itself is incapable of exciting kink 
production; we restrict the amplitude to $\chi_0 \leq 0.38$. 

  The dependence of the kink density on the dissipation coefficient is 
depicted in Fig.6. We emphasize that our results are valid only for very
low dissipation coefficients. For large dissipation coefficients, the 
field oscillations are considerably suppressed leading to a suppression 
in kink densities as evident from the plots of Fig.6.

  The kink densities are also crucially dependent on the choice of frequency
of the oscillating background field. We have found that large amplitude field
oscillations are induced, leading to kink-antikink production, only for 
frequencies lying in the range $0.3 \le \omega \le 2.5$. 
For frequencies beyond this window, no significant enhancement of kink 
densities compared to their thermal equilibrium value is observed. As is
evident from Fig.7, there exists an optimum value of frequency (all other
parameters remaining fixed) for which kink density is maximized. This 
optimum value also depends on the temperature of the heat bath ($T$), 
damping coefficient ($\eta$) and the amplitude of oscillations ($\chi_0$).

\section{Discussions and Conclusion}

  In this paper, we have investigated a novel phenomenon in a 
$(1+1)$ dimensional field theory admitting topological solitons called 
kinks. We investigated the production of kink-antikink pairs when the system 
is subject to the twin effects of noise and  periodic driving via its 
coupling to an oscillating but spatially homogeneous background 
field. For a certain range of frequencies of the oscillating background,
there occurs considerable enhancement in the densities of kinks compared
to their thermal equilibrium values. We also studied the effect on kink
density of parameters like the damping coefficient, the temperature of the 
heat bath and the amplitude of oscillations of $\chi$. Our results are 
particularly sensitive to the value of the dissipation coefficient and 
the enhancement in kink density compared to their thermal equilibrium value 
is observed only for low dissipation coefficients. Kink-antikink pair
production is observed even though the system remains in the broken
phase throughout the course of the simulations. This further demonstrates 
that topological defects need not be produced only during symmetry breaking 
phase transitions, as pointed out earlier \cite{ray}. 

  At this stage it is tempting to compare our results with the intriguing
phenomenon of stochastic resonance (SR) which has been extensively investigated
in the literature on non-linear dynamical systems \cite{stch}. The 
characteristic feature of SR is that the an increase in the noise strength can 
sometimes lead to more coherent behavior when the non-linear dynamical 
system is also subject to a periodic driving force. In particular, by tuning
the noise strength, a significant improvement in the signal-to-noise 
ratio (also manifest through peaks in the noise averaged power spectrum)
is achieved. The study of stochastic resonance for spatially extended 
systems has been carried out for Ginzburg-Landau type field theories, 
albeit restricted to the over-damped regime \cite{march,benzi}.
There it was found that an appropriate choice of the frequency of the periodic
driving obtained by matching the thermal activation time-scale (given 
by the inverse of Kramers rate) to half the period of the modulating 
background, can result in periodically synchronized behavior of the 
mean field about $\phi =0$ (see Fig.2 of Ref. \cite{benzi}). In our case
however, no synchronization of the mean field is observed, rather it is
found to decay to zero from its initial value in one of the vacuum states,
and thereafter keep on oscillating erratically about the zero field value.
As has been demonstrated (see Fig.3), this behavior can be attributed 
to kink-antikink production which occurs inspite of the fact that the 
system remains in the broken phase. This comparison makes it clear that 
the phenomenon we have discussed is quite distinct from that of 
stochastic resonance. The investigation of the phenomenon of stochastic
resonance in underdamped system is currently in progress \cite{ssg2}.

 There is much that needs to be investigated. Apart from the study of the
phenomenon of SR in underdamped systems, a theoretical understanding of 
the phenomenon discussed here is required. In particular, an analytical
derivation of the frequency window required for enhanced kink-antikink
production would be both interesting and useful. Moreover, the issue of 
lattice spacing dependence of results of non-equilibrium dynamical systems 
requires clarification. Also, a study of kink production in the Sine-Gordon 
model coupled to an oscillating background would be interesting. We plan to 
address these issues in a future work.

\section{Acknowledgements}

 We would like to thank A.M. Srivastava for enlightening discussions and
encouragement. SS would also like to thank Marcelo Gleiser, F.C. Khanna 
and Surujhdeo Seunarine for useful discussions. The work of SS was funded 
in part by NSERC, Canada. 
 


\newpage

\vskip -0.25in
\begin{figure}[h]
\begin{center}
\leavevmode
\epsfysize=18truecm \vbox{\epsfbox{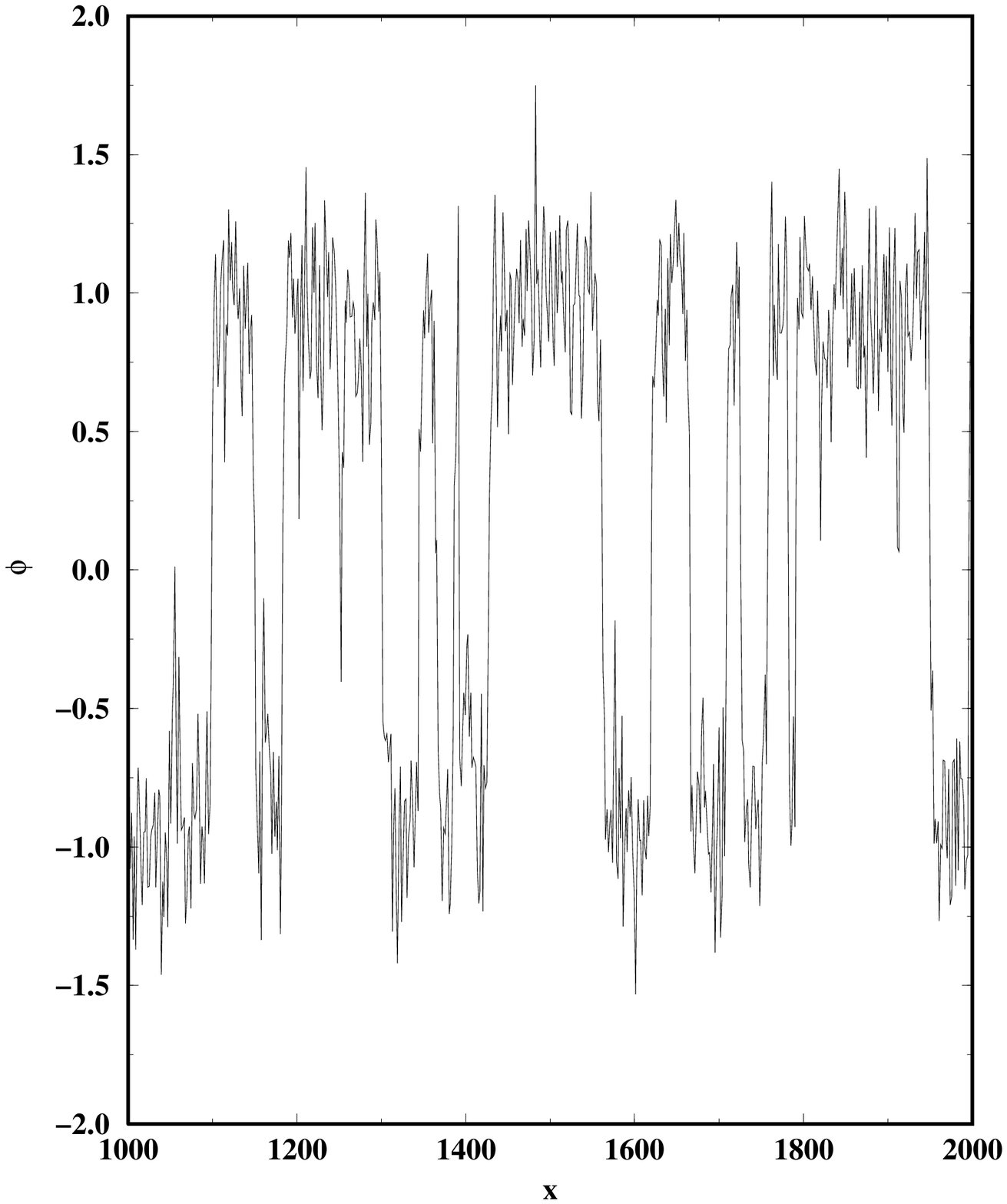}}
\end{center}
\vskip +0.2in
\caption{The field configuration at t=3000, for a portion of the lattice 
showing kink-antikink pairs.The parameter values are $T=0.12, \eta=0.01,
\chi_{0}=0.34, \omega=1.10$.} 
\label{Fig.1}
\end{figure}

\newpage

\vskip -0.35in
\begin{figure}[h]
\begin{center}
\leavevmode
\vskip -0.35in
\epsfysize=20truecm \vbox{\epsfbox{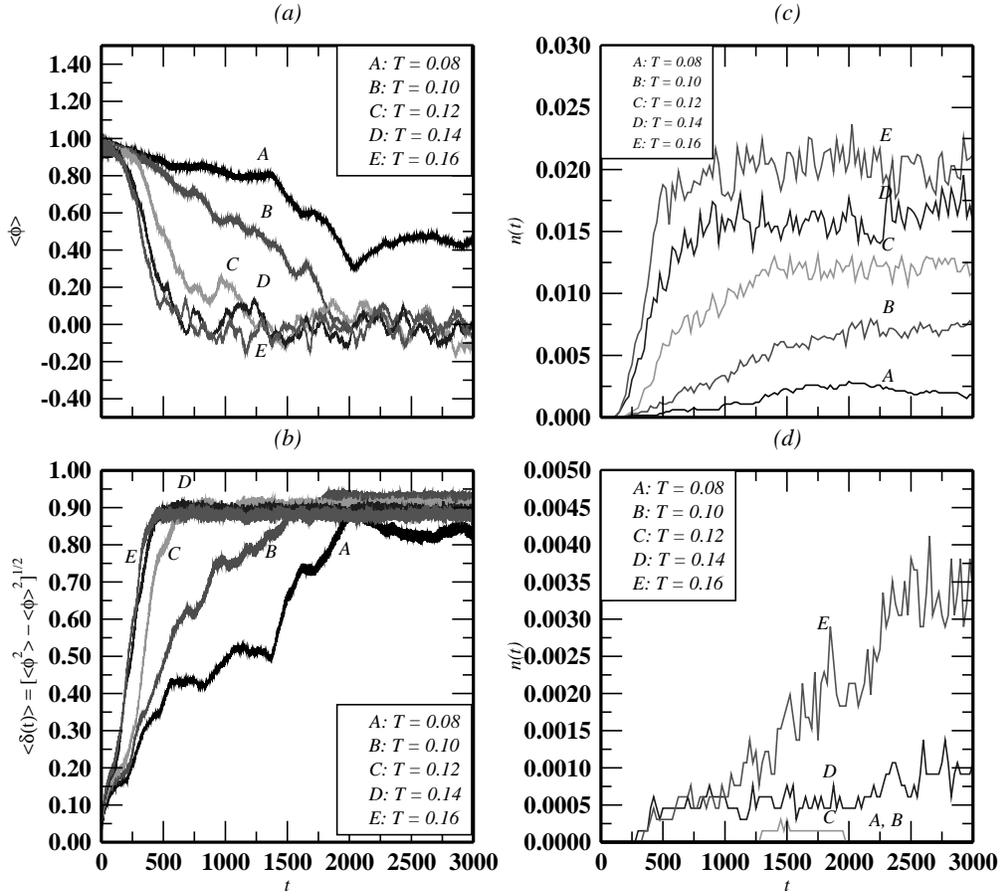}}
\end{center}
\vskip +0.2in
\caption{The time evolution of the (a) mean field, (b) fluctuations,
(c)kink-antikink density (d) kink-antikink density in the absence
of the oscillating background ($g^2=0$); for five different bath 
temperatures. The plots labeled A,B,C,D,E correspond to temperatures 
$T=0.08,0.1,0.12,0.14,0.16$ respectively. Other parameter values which are 
kept fixed are $\eta=0.01$, $\omega=1.10$, $\chi_0=0.34$. Note the difference
in scales on the y-axis of 2(c) and 2(d).}
\label{Fig.2}
\end{figure}

\newpage

\begin{figure}[h]
\begin{center}
\leavevmode
\vskip -4.50in
\epsfysize=18truecm \vbox{\epsfbox{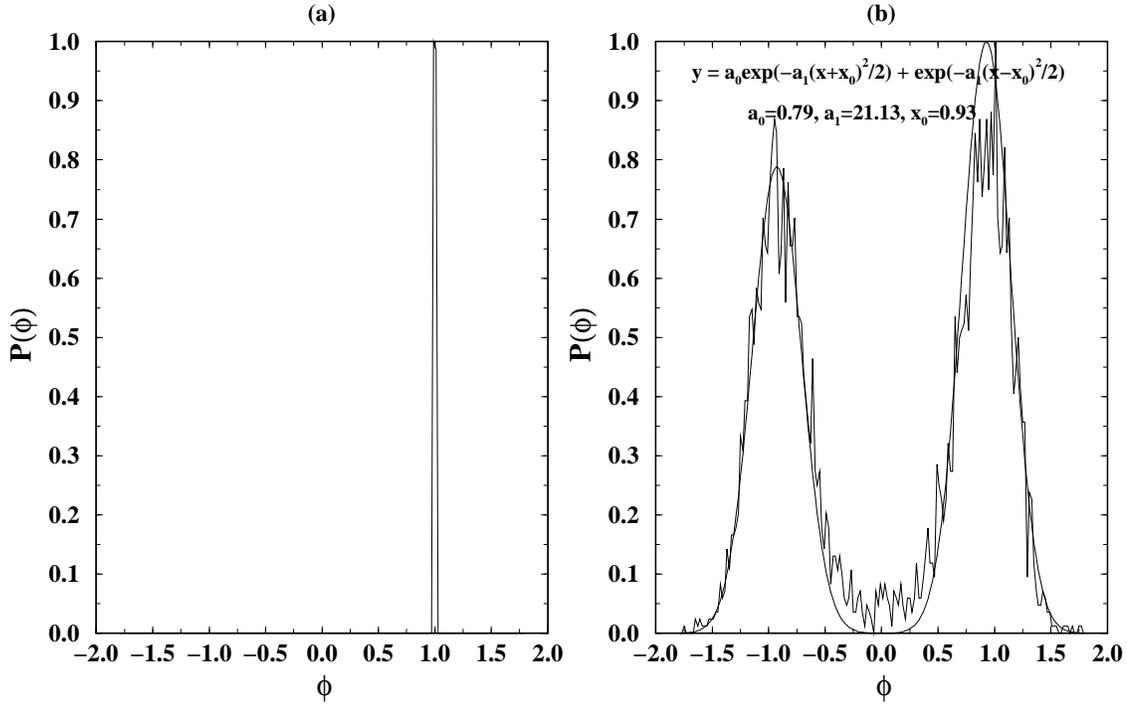}}
\end{center}
\vskip +0.2in
\caption{(a) Probability distribution of $\Phi$ at (a) $t$ = 0 
(b) $t$=3000.The solid line in (b) is an asymmetric double-gaussian fit of 
the data given by the function $a_{0}*exp[-(a_{1}/2)*(x+x_{0})^2] + 
exp[-(a_{1}/2)*(x-x_{0})^2]$, where $a_{0}=0.79, a_{1}=21.13, x_{0}=0.94$. 
The large twin peaks around $\Phi=\pm 1$, even at late times, is a clear 
indication that the symmetry remains spontaneously broken.} 
\label{Fig.3}
\end{figure}

\newpage

\begin{figure}[h]
\begin{center}
\leavevmode
\vskip -3.5in
\epsfysize=18truecm \vbox{\epsfbox{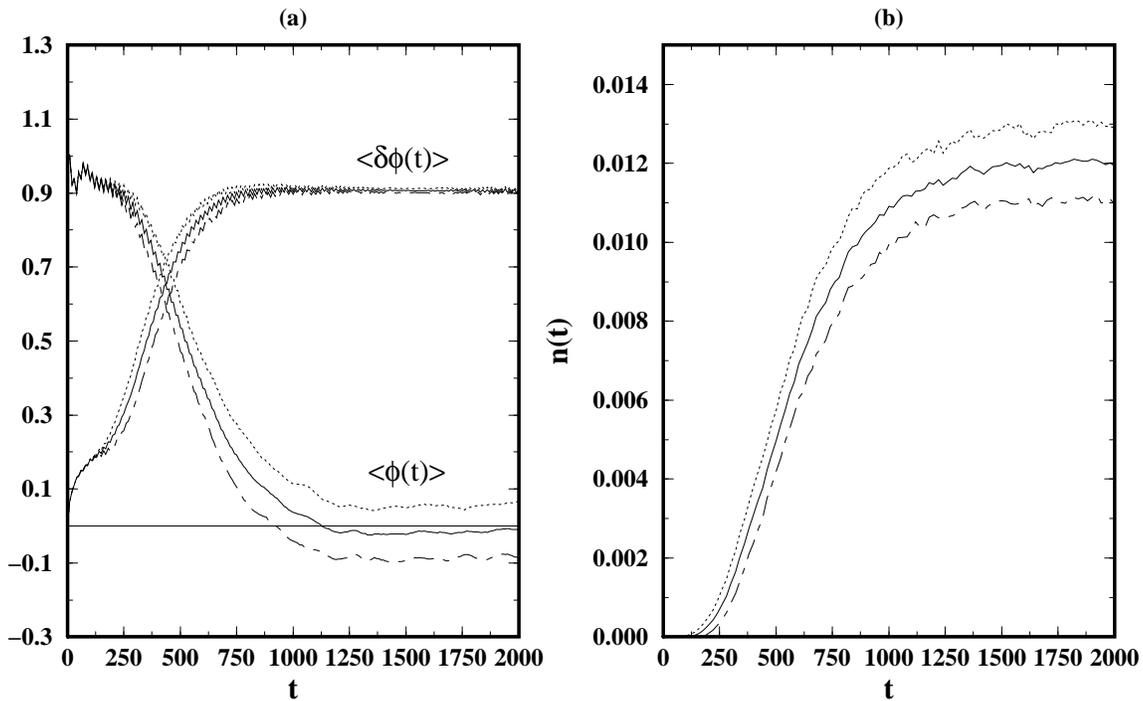}}
\end{center}
\vskip +0.2in
\caption{Time-evolution of the noise averaged values of the (a)mean field 
and fluctuations,(b) kink-antikink density, together with the $\pm 1\sigma$ 
error. The solid line indicates the noise-averaged mean while the dashed 
lines above and below the mean correspond to $+ 1\sigma$ and $- 1\sigma$ 
deviation from the noise averaged value. Noise average has been carried out 
over 100 different noise realizations.} 
\label{Fig.4}
\end{figure}

\newpage

\vskip -0.35in
\begin{figure}[h]
\begin{center}
\leavevmode
\epsfysize=20truecm \vbox{\epsfbox{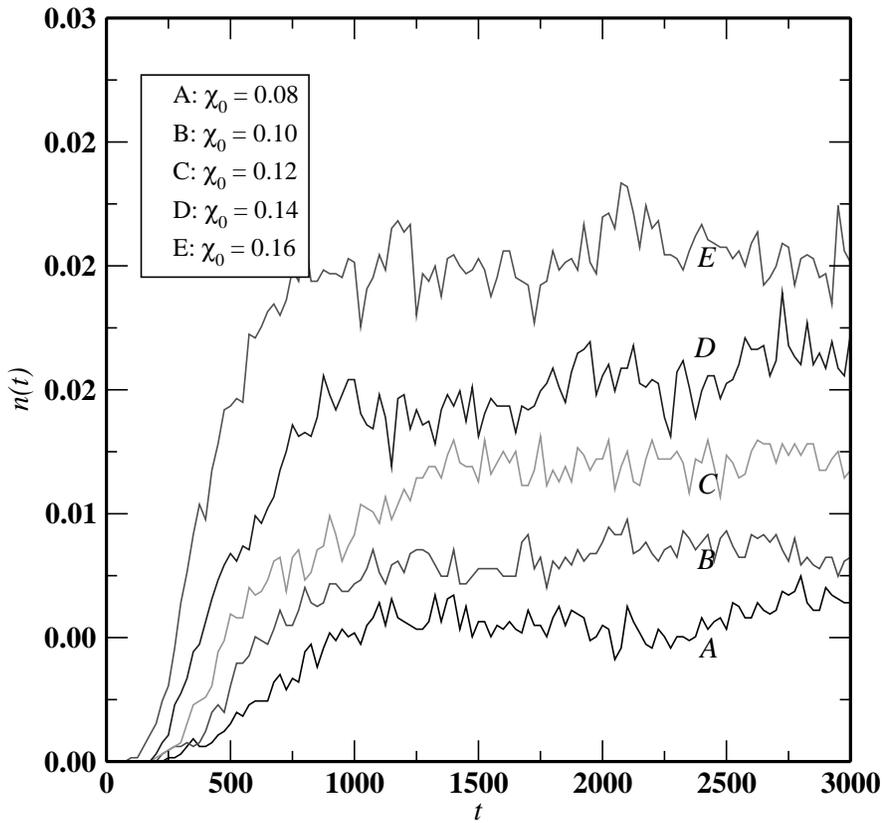}}
\end{center}
\vskip +0.2in
\caption{The variation of the kink-antikink density with the amplitude
$\chi_{0}$ of the oscillating background field. Other parameter values are
kept fixed at $T=0.12, \eta=0.01, \omega=1.10$.}
\label{Fig.5}
\end{figure}

\newpage

\vskip -0.35in
\begin{figure}[h]
\begin{center}
\leavevmode
\epsfysize=20truecm \vbox{\epsfbox{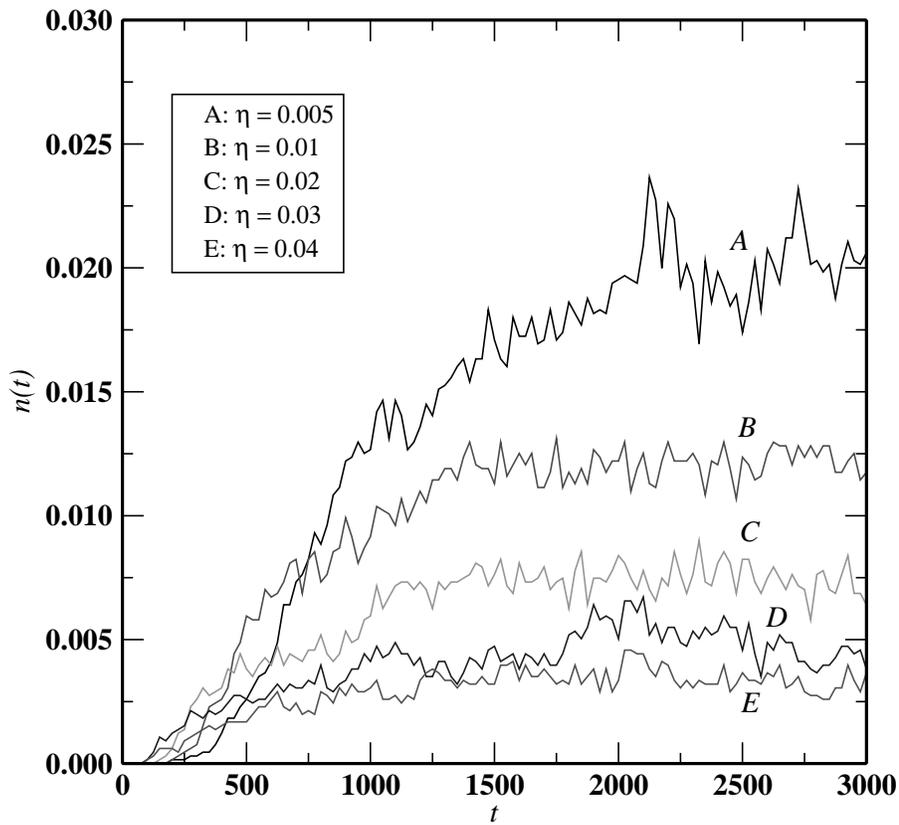}}
\end{center}
\vskip +0.2in
\caption{The variation of the kink-antikink density with the dissipation 
coefficient $\eta$. Other parameter values are kept fixed at $T=0.12, 
\chi_0=0.34, \omega=1.10$.}
\label{Fig.6}
\end{figure}

\newpage

\vskip -0.35in
\begin{figure}[h]
\begin{center}
\leavevmode
\epsfysize=20truecm \vbox{\epsfbox{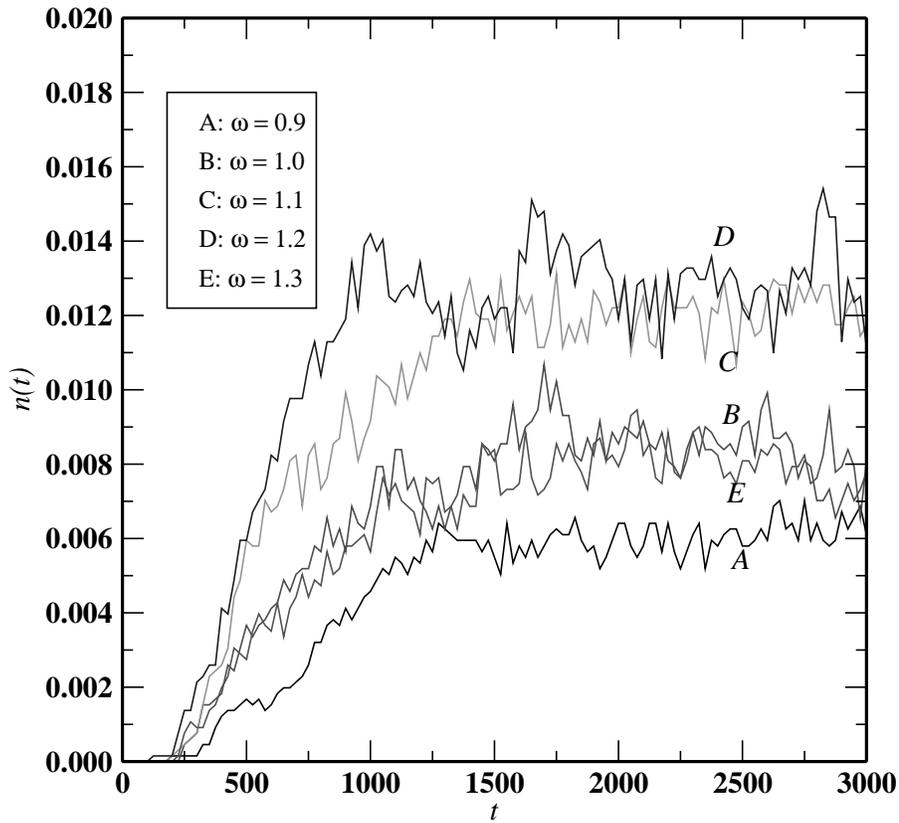}}
\end{center}
\vskip +0.2in
\caption{The variation of the kink-antikink density with the frequency 
$\omega$ of the oscillating background. Other parameter values are kept 
fixed at $T=0.12, \chi_0=0.34, \eta=0.01$.}
\label{Fig.7}
\end{figure}

\end{document}